# Electronic structure of porphyrin-based metal-organic frameworks and their suitability for solar fuel production photocatalysis

Said Hamad,[a]* Norge C. Hernandez,[b] Alex G. Aziz,[c] A. Rabdel Ruiz-Salvador,[a] Sofia Calero,[a] and Ricardo Grau-Crespo[c]†

Metal-organic frameworks (MOFs) can be exceptionally good catalytic materials thanks to the presence of active metal centres and a porous structure that is advantageous for molecular adsorption and confinement. We present here a first-principles investigation of the electronic structure of a family of MOFs based on porphyrins connected through phenyl-carboxyl ligands and AlOH species, in order to assess their suitability for the photocatalysis of fuel production reactions using sunlight. We consider structures with protonated porphyrins and those with the protons exchanged with late 3d metal cations ($Fe^{2+}$, $Co^{2+}$, $Ni^{2+}$, $Cu^{2+}$, $Zn^{2+}$), a process that we find to be thermodynamically favorable from aqueous solution for all these metals. Our band structure calculations, based on an accurate screened hybrid functional, reveal that the bandgaps are in a favorable range (2.0 to 2.6 eV) for efficient adsorption of solar light. Furthermore, by approximating the vacuum level to the pore center potential, we provide the alignment of the MOFs' band edges with the redox potentials for water splitting and carbon dioxide reduction, and show that the structures studied here have band edges positions suitable for these reactions at neutral pH.

[a.] Departamento de Sistemas Físicos, Químicos y Naturales, Universidad Pablo de Olavide, Carretera de Utrera km. 1, 41013 Sevilla, Spain. *E-mail address: said@upo.es.
[b.] Departamento de Física Aplicada I, Escuela Técnica Superior de Ingeniería Informática, Avenida Reina Mercedes, Universidad de Sevilla, 41012 Sevilla, Spain.
[c.] Department of Chemistry, University of Reading, Whiteknights, Reading RG6 6AD, United Kingdom.
* E-mail address: said@upo.es.
† E-mail address: r.grau-crespo@reading.ac.uk.

## INTRODUCTION

The development of cheap, efficient techniques to carry out the photocatalytic splitting of water would permit the generation of a clean fuel (hydrogen) at low negative environmental impact.[1-3] Of similar interest is the photocatalytic reduction of $CO_2$, which would allow the synthetic production of carbon-containing fuels (e.g. methanol)[4,5] and simultaneously contribute to recycle $CO_2$ from the environment. It is clear that the development of efficient technologies to carry out these energetically up-hill reactions using solar energy would be greatly beneficial, and therefore, many research efforts are being devoted to it, in particular to the search for adequate photocatalysts.[6-8] This research was initially focused on traditional inorganic semiconductors, such as $TiO_2$ and CdS[3, 9-12], but it has now extended to a wider class of materials, including nanostructures such as fullerenes, nanotubes and graphene-like 2D solids.[13-17] On the other hand, water splitting and $CO_2$ reduction are at the core of natural photosynthetic reactions,[18] so the study of the related natural processes can help finding artificial routes for these reactions.[3, 19-21] Bioinspired molecular photocatalysts have been largely studied in recent years.[22-24]

Besides Mn-complexes that are close to natural photosynthesis reactions,[23, 25, 26] porphyrins have also been identified as active molecular centres for artificial photosynthesis.[27, 28] One drawback of molecular systems from a practical point of view is their recyclability, as the separation of the catalysts from liquid media is very difficult. To overcome this problem, one attractive route is their immobilization in solid hosts.[29, 30] Metal-organic frameworks (MOFs) have appeared as promising hosts,[31-34] where catalytic centres can be encapsulated[35-39] or moreover be part of the constituents of the materials.[39-42] After pioneering work by Suslick and co-workers,[43] a number of porphyrin-based MOFs have been reported,[44-47] including some with photocatalytic properties.[44-46]

The choice of MOFs as catalysts is also stimulated by their diverse porous architectures,[48, 49] which confer them exceptional molecular adsorption properties.[49-51] In these systems the adsorbed reactant molecules can access the active sites embedded in a confinement field that favours the catalytic reactions[52]. Furthermore, it is possible to tune the adsorption properties of MOFs, by changing factors such as topology, metal composition or the nature of the ligand.[53, 54] However, the impact of these modifications on the electronic structure of MOFs has not been widely studied. Gascon et al.[55] showed that by modifying the linker properties the overall bandgap of MOF-5 can be lowered. The dependence of the optical properties of a nanotubular MOF on the adsorbed guest molecules can be exploited in molecular sensing applications.[56] The magnetic and optical properties can also be tuned by the composition of the metal centres.[57, 58] The catalytic behaviour of nanoporous solids strongly depends on both the structural and electronic features of the materials, so a rational design of these materials would make them very useful in catalytic applications.

In contrast to semiconductor photocatalysts, in terms of their electronic behaviour, MOFs can be regarded as molecular-like catalysts.[59] Fateeva and co-workers[46] have showed that the optical and photocatalytic properties of a MOF containing porphyrins connected through phenyl-carboxyl ligands and AlOH species (called Al-PMOF here) are primarily determined by the porphyrin linker. For Al-PMOF these authors observed a strong absorption band at 415 nm (2.99 eV) and four lower energy bands, which are characteristic of the free porphyrin molecule in solution. The photocatalytic properties of porphyrin molecules can in principle be modified by the presence of metal (M) cations within the ring.[60] For example, metalation of Al-PMOF with Zn has been investigated to improve the efficiency of photocatalytic water splitting,[46] while $CO_2$ conversion is improved by Cu incorporation.[61]

It is therefore relevant to study the electronic structure of metallated Al-PMOF materials in order to understand the effect of the porphyrin metal on photocatalytic reactions. In this paper we consider the incorporation of Fe, Co, Ni, Cu or Zn in the centre of the porphyrins in Al-PMOF, in comparison also to the di-protonated Al-PMOF. For this purpose, we use Density Functional Theory (DFT) to calculate the electronic bandgap, and the absolute positions of the band edges, as these are important factors in determining the suitability and efficiency of a material as photocatalysts.

## COMPUTATIONAL METHODS

In our calculations the structures were represented by a primitive rhombohedral cell, which contains half the number of atoms as the conventional orthorhombic cell shown in Figure 1 (space group Cmmm)[46], leading to only one porphyrin per cell. Inside the porphyrin, bonding to the N atoms, we place either two hydrogens, or a late 3$d$ transition metal cation ($Fe^{2+}$, $Co^{2+}$, $Ni^{2+}$, $Cu^{2+}$ or $Zn^{2+}$).

We performed spin-polarized quantum-mechanical calculations using density functional theory (DFT) as implemented in the Vienna *Ab-initio* Simulation Package (VASP).[62-65] Geometry optimizations were performed using the generalized gradient approximation (GGA) with the Perdew-Burke-Ernzerhof functional (PBE) functional.[66, 67] During relaxation, forces on atoms were minimized until they were all less than 0.01 eV/Å.

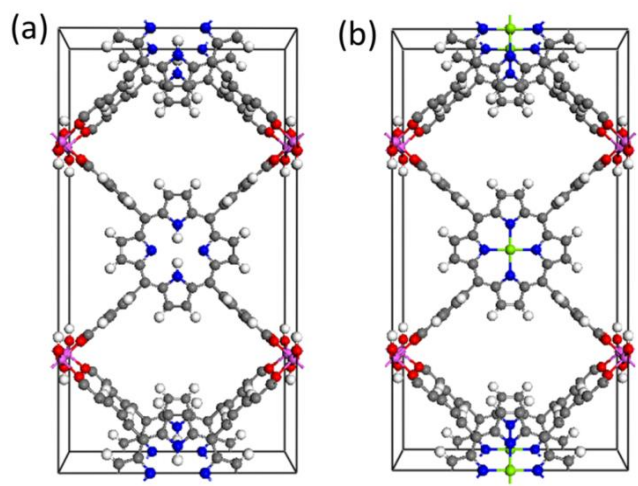

Figure 1. Perspective view of the porphyrin-based MOF investigated in this study in a) the protonated case, and b) the metal-substituted case. Color code: Gray=Carbon, White=Hydrogen, Red=Oxygen, Blue=Nitrogen, Magenta=Aluminium, Green=Transition Metal. The unit cell shown here is C-centered orthorhombic, but a primitive cell, with half the number of atoms of the orthorhombic cell, was used in our calculations.

For each of the metal ions, we considered all possible spin states. For $Fe^{2+}$ ($d^6$ ion) there can be 0, 2 or 4 unpaired electrons, which correspond to low-spin (LS), intermediate-spin (IS) and high-spin (HS) configurations. For $Co^{2+}$ ($d^7$) there can be either 1 (LS) or 3 (HS) unpaired electrons, whereas for $Ni^{2+}$ ($d^8$) there can be either 0 (LS) or 2 (HS) unpaired electrons. Finally, there is one possible spin state for $Cu^{2+}$ ($d^9$) with one unpaired electron, whereas $Zn^{2+}$ is not spin-polarised. In each calculation, we constrained the difference in the number of up- and down-spin electrons to the corresponding integer number given above. Because there is only one metal atom per cell, all our spin-polarised calculations correspond to ferromagnetic configurations. Considering the relatively large distance between magnetic metal ions (ca. 6.7 Å in the direction perpendicular to the porphyrin plane) and the absence of common ligands, it can be expected that the strength of the magnetic coupling will be very small and that the actual structure is paramagnetically disordered at room temperature. However, the weak magnetic coupling implies that there must be very small energy differences associated with magnetic ordering in these structures, and therefore the simulation results should be independent on the magnetic order assumed.

In order to obtain accurate electronic structures we carried out single-point calculations at the most favourable spin state for each composition, using the screened hybrid functional of Heyd, Scuseria and Ernzerhof (HSE06),[68, 69] which generally provides bandgaps in closer agreement with experiment than those from GGA functionals.[70] Although the HSE06 calculations were single-point only, based on the PBE-optimised structures, test calculations showed that the final electronic structures were largely insensitive to the small geometric variations introduced by re-optimising the structures at more sophisticated levels of calculation, e.g. adding corrections to improve the description of $d$ orbitals and of van der Waals interactions (see Supplementary Information).

The projector augmented wave (PAW) method[71, 72] was used to describe the frozen core electrons and their interaction with the valence electrons, *i.e.* those in level $4d$ for Fe, Co, Ni, Cu and Zn, and $2s2p$ for C, N and O. The kinetic energy cutoff for the plane-wave basis set expansion was set at 600 eV. A $\Gamma$-centred grid of $k$-points was used for integrations in the reciprocal space, where the smallest allowed spacing between $k$-points was set at 0.5 Å$^{-1}$, giving rise to 3 irreducible points in the Brillouin zone corresponding to the primitive cell.

As in other periodic DFT codes, the band energies in VASP are given with respect to an internal energy reference (the average potential in the crystal). In order to align the band energies with the vacuum scale, it is necessary to evaluate the electrostatic potential in the vacuum region represented by an empty space within the simulation cell. In the present study we follow the methodology recently proposed by Butler *et al.*[73] to calculate the vacuum level in MOF structures, which consists of evaluating the average potential within a small sphere (radius of 2 Å) at different positions in the pore. By finding the point that is farthest apart from the framework atoms (the pore "centre"), where the potential is locally flat (no electric field), we can obtain a good approximation to the vacuum level. In Ref. 73, this procedure led to MOF ionization potentials in good agreement with experiment. A Python code provided by these authors was employed in our calculations to obtain the average potentials.[74]

## RESULTS AND DISCUSSION

### Crystal structures

We first examine the crystal structures resulting from the substitutions of different cations in the porphyrin. Upon replacement of the two H atoms by the transition metal atoms, only small variations of cell parameters are observed (actions between 0.1% and 0.6%.

Table 1), with overall cell volume contractions between 0.1% and 0.6%.

Table 1. Calculated lattice parameters, cell volume and the two perpendicular N-N distances inside the porphyrin. All structures adopt the orthorhombic space group *Cmmm* (65), where α=β=γ=90°. Available experimental values at room temperature are given in parenthesis.

| Cation | $a$ (Å) | $b$ (Å) | $c$ (Å) | $V$ (Å3) | $d$ [N-N] (Å)* |
|---|---|---|---|---|---|
| 2H$^+$ | 32.196 (31.967)† | 6.722 (6.6089) | 16.964 (16.876) | 3671.6 (3565.3) | 4.06/4.22 - |
| Fe$^{2+}$ | 32.072 | 6.720 | 16.957 | 3654.8 | 3.96/3.96 |
| Co$^{2+}$ | 31.968 | 6.720 | 16.989 | 3649.3 | 3.92/3.93 |
| Ni$^{2+}$ | 31.996 | 6.723 | 16.964 | 3649.2 | 3.91/3.91 |
| Cu$^{2+}$ | 32.079 | 6.717 | 16.976 | 3658.0 | 4.02/4.03 |
| Zn$^{2+}$ | 32.061 (31.861)† | 6.726 (6.601) | 17.004 (16.895) | 3667.0 (3553.1) | 4.09/4.10 - |

* In the metal-substituted porphyrins, the reported d(N-N) corresponds to twice the cation-N distance. †Experimental data at room temperature for the taken from Section 8.5 of the supplementary information of Ref 46.

The metal-nitrogen interaction seems to be the main factor controlling the small changes of the cell volume. The calculations systematically overestimate all the cell parameters in comparison with experiment in Ref. 46 by an average of ~1%. However, the calculations are able to reproduce well the small variations observed experimentally in the cell parameters from the protonated to the Zn-substituted structure: a small contraction of the $a$ parameter, a small expansion of the $c$ parameter, while the $b$ parameter remains roughly the same. Overall, a small contraction of the cell volume by is observed in both the experiment (0.31%) and the calculations (0.13%) upon Zn substitution.

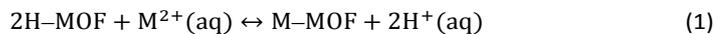
To appear in Journal of Materials Chemistry A (Accepted on 9th October 2015)**Thermodynamics of metal substitutions from aqueous solution**

In order to discuss the stability of the material with the different metal cation ($M^{2+}$) substitutions, we have calculated the enthalpies and free energies for the process of cation exchange with aqueous solution, according to the following reaction:

$$2H\text{–MOF} + M^{2+}(aq) \leftrightarrow M\text{–MOF} + 2H^+(aq) \quad (1)$$

using a mixed theoretical-empirical approach for the treatment of the aqueous cations. For each cation, the enthalpy in aqueous solution is approximated using the following quantities: a) the DFT energy of the neutral atom in the gas phase, calculated with VASP at the experimental spin groundstate, using a large supercell and the same functional and precision parameters as in the MOF calculations; b) the sum of the experimental first and second ionization energies,[75, 76] which is the energy needed to ionize the neutral atom to a $M^{2+}$ cation (of course, only the first ionization energy is used in the case of hydrogen); and c) the experimental hydration enthalpy,[77] which is the enthalpy change in the process of moving the $M^{2+}$ cation from the gas phase to aqueous solution. The addition of these three contributions gives the enthalpy of the aqueous cation, which will be used in the calculation of the enthalpy change of reaction (1), $\Delta H$. On the other hand, the reaction free energy $\Delta G$ can be estimated as:

$$\Delta G = \Delta G_0 + k_B T \ln\left(\frac{[H^+]^2}{[M^{2+}]}\right) \quad (2)$$

where $\Delta G_0$ is calculated using the same procedure as for $\Delta H$, but employing the experimental hydration free energies of the cation (instead of the hydration enthalpy), i.e., considering the entropy contribution to hydration.[78] The second term, where $k_B$ is Boltzmann's constant, takes into account the effect of the relative concentrations of the cations in aqueous solution. The obtained values for $\Delta H$ and $\Delta G_0$ are reported in Table 2. As shown in Figure 2, the variation with cation concentration and pH is relatively weak. Increasing the solution pH (i.e. decreasing the proton concentration) or increasing the metal concentration in solution, makes the exchange reaction slightly more favourable.

The calculated free energies for reaction (1) are negative for all the cations at any reasonable concentration of the cations in aqueous solution and pH values, which indicates that immersing the 2H-MOF in an aqueous solution of $M^{2+}$ cations would lead to spontaneous exchange with the cations from the solution substituting the protons at the centre of the porphyrin rings. The exchange process is partly driven by entropy effects. Indeed, the enthalpy change, $\Delta H$, of the exchange process for $Fe^{2+}$ is positive, and it is the introduction of the hydration entropy effect through the hydration free energy what makes the exchange spontaneous. This analysis suggests that the substitution of these metals in the porphyrin-based MOF should be straightforward in experiment. This is agreement with the observed experimentally easy introduction of $Cu^{2+}$ in Al-PMOF.[61] The preference for particular spin states, which can be seen from Table 2, will be discussed below in terms of the electronic structure.

Table 2. Enthalpy and free energy change of the process of exchanging the two protons at the centre of the porphyrin by a $M^{2+}$ cation, for the various possible spin moments (μ). LS, low-spin. IS, intermediate-spin. HS, high-spin. Bold font is used to highlight the most stable spin state for each composition.

| Cation | Spin State | μ (μ$_B$) | $\Delta H$ (eV) | $\Delta G_0$ (eV) |
|---|---|---|---|---|
| $Fe^{2+}$ | LS | 0 | 0.61 | -0.82 |
| | **IS** | **2** | **0.20** | **-1.23** |
| | HS | 4 | 0.91 | -0.52 |
| $Co^{2+}$ | **LS** | **1** | **-2.16** | **-3.28** |
| | HS | 3 | -1.34 | -2.46 |
| $Ni^{2+}$ | **LS** | **0** | **-3.05** | **-4.60** |
| | HS | 2 | -1.93 | -3.48 |
| $Cu^{2+}$ | - | 1 | **-5.86** | **-6.86** |
| $Zn^{2+}$ | - | 0 | **-4.52** | **-5.58** |

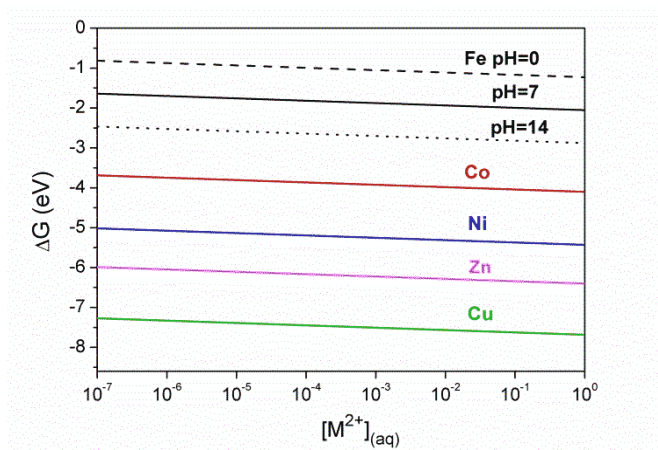

Figure 2. Free energy of the cation exchange reaction (as shown in equation 1), as a function of type of cation, cation concentration in solution and pH (at room temperature). All solid lines correspond to values at pH=7. For Fe we also show the two lines corresponding to pH=0 and pH=14. For the other cations the variation with pH is the same as in the case of Fe, so for clarity reasons only the results at neutral pH are shown.

**Electronic structure**

The total electronic density of states (DOS) and their projections on the 3$d$ orbitals of the metals are shown in Figure 3. All structures are semiconductors, with bandgaps in the range between 2.0 and 2.6 eV. A bandgap of around 2 eV is generally considered to be ideal for single-semiconductor water splitting photocatalysis.[79] Anatase $TiO_2$, one of the most widely investigated photocatalysts for water splitting, has a bandgap of 3.2 eV,[80] which has to be engineered via doping in order to favour the absorption of the visible component of the solar spectrum.[81, 82] MOFs with adequate band alignment for photocatalysis were reported by Butler et al.,[73] although the calculated bandgaps were wider (above 3 eV in all cases).

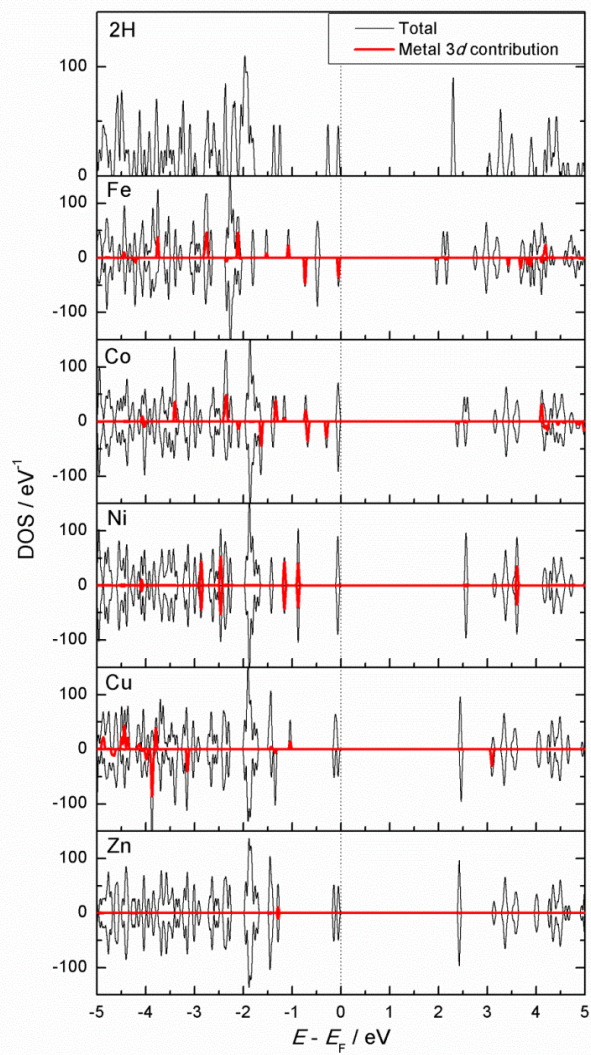

Figure 3. Electronic density of states (DOS) of the protonated and metal-substituted porphyrin MOFs, as obtained using the screened hybrid functional HSE06.

In order to understand the position in the DOS of the metal 3$d$ contributions we need to refer to their splitting due to the porphyrin ligand field. Although the position of porphyrin as a ligand in the spectrochemical series is considered to be ambiguous, it is known that in metallo-porphyrins with square planar coordination of divalent cations, porphyrin tends to act as a strong field ligand.[83] That means that the $d_{x^2-y^2}$ level is very high up in energy compared with the others, as shown schematically in Figure 4. But in this MOF, the degeneracy of the two lower $d$ levels is broken, because the crystal does not have the 4-fold rotation axis that would be present in the isolated gas-phase porphyrin. The distortion could be then referred to as rectangular distortion, to indicate the lowering of the symmetry to a 2-fold rotation axis.

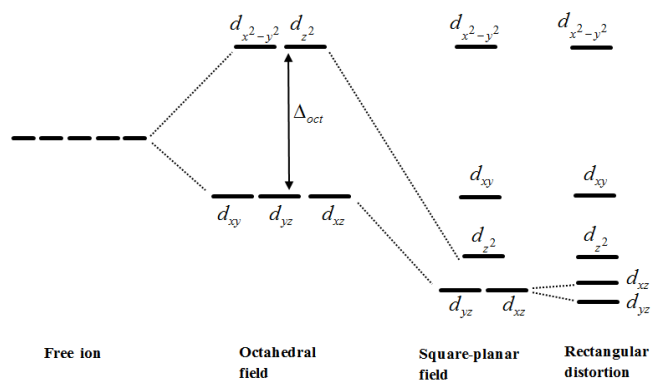

Figure 4. Scheme of the splitting of *d* orbitals in a square planar field, and the effect of a rectangular distortion.

Making use of the splitting of levels shown in Figure 4, we can explain the spin state of the MOF with each cation:

a) **Fe$^{2+}$ ($d^6$)**. The most favourable configuration is an intermediate-spin state with magnetic moment μ(=2$S$)=2 per Fe$^{2+}$ cation. The high-spin state (μ=4) is forbidden by the strong splitting, since it would require the promotion of an electron to the $d_{x^2-y^2}$ level. But the separation between the rest of the *d* levels is not strong enough to lead to low-spin state. Following Hund's rule, the favourable state is therefore the intermediate-spin one (μ=2), with the $d_{xy}$ and $d_{xz}$ levels doubly occupied and $d_{xy}$ and $d_{z^2}$ singly occupied. The intermediate-spin groundstate of Fe$^{2+}$ in porphyrin is consistent with previous theoretical results for porphyrin molecules.[84]
b)       **Co$^{2+}$ ($d^7$)**. This cation has one electron more than Fe$^{2+}$. For that reason, the $d_{z^2}$ level is doubly occupied and the only unpaired electron is in the $d_{xy}$ level, leading to a low-spin state (μ=1).
c)       **Ni$^{2+}$ ($d^8$)**. In this case the four low energy levels are doubly occupied in the low-spin state (μ=0). The high-spin state (μ=2) would require the promotion of one electron from the $d_{xy}$ level to the $d_{x^2-y^2}$ level, making it energetically unfavourable.
d) **Cu$^{2+}$ ($d^9$)**. There is only one possible spin state for this cation (μ=1), since the four low energy levels are filled and one electron occupies the $d_{x^2-y^2}$ level.
e) **Zn$^{2+}$ ($d^{10}$)**. In this case, the five *d* orbitals are filled, so μ=0.

Regarding the bandgap, we can see in Figure 3 that all materials (except Fe-Al-PMOF) have similar values of bandgap, between 2.3 eV and 2.6 eV. The projection of the DOS on individual atoms and orbitals (not shown in the figure) reveals that the HOMO is associated with the *p* orbitals of the N and C atoms of the porphyrin moieties, while the LUMO is associated with the *p* orbitals of the C and O atoms of the carboxyl ligands. In the case of Fe-Al-PMOF, the Fe atoms introduce the $d_{xy}$ levels into this otherwise unoccupied region (close to the valence band), leading to a decrease in bandgap to 2.02 eV. Note that the DOS plots in Figure 3 are aligned with respect to the Fermi level, which means that the introduction of a peak in the band gap region shifts all the peaks to the left in the case of Fe. The bandgaps of the materials clearly make them good candidates to carry out photocatalytic reactions, since they would be able to absorb most of the solar radiation. However, in order to assess the ability of a material to perform photocatalytic water splitting or carbon dioxide reduction, we also need to investigate if the semiconductors exhibit the correct alignment of the bands with respect to the half-reaction potentials, which we do in the following section.

**Band edge positions with respect to electrode potentials**

The alignment of the bands with the vacuum level allows us to explore the thermodynamic feasibility of the photocatalytic processes. A single-semiconductor photocatalyst requires certain important characteristics in its electronic structure. For example, for the water splitting reaction, the positions of the conduction and valence band edges should straddle the redox potentials for water photolysis,[79, 85, 86] i.e. the valence band edge should be below the energy of the oxygen evolution reaction (OER):

$$H_2O \leftrightarrow 2H^+_{(aq)} + \tfrac{1}{2}O_{2(g)} + 2e^-, \qquad (3)$$

and the conduction band edge should be above the energy corresponding to the hydrogen evolution reaction (HER):

$$2H^+_{(aq)} + 2e^- \leftrightarrow H_{2(g)}. \tag{4}$$

The energy scale has the opposite sign of the potential scale, so lower energy means higher potential, and *vice versa*. The bandgap must therefore be wider than 1.23 eV (difference between the HER and the OER levels). After loss mechanisms are accounted for, a bandgap of 2 eV or more is generally considered as necessary,[79] but the bandgap should not be too wide, in order to allow the adsorption of photons from the visible part of solar radiation. It is known that, in the vacuum scale and at pH=0, the HER level is located at -4.44 eV, and the OER level is located at -5.67 eV.[87] At temperature $T$ and pH > 0, these energy levels are shifted up by $pH \times (k_B T \times ln10)$. By referencing the electronic levels in our semiconductor solids with respect to the vacuum level (taken here as the electron potential at the centre of the largest pore), we can assess whether the band edges of the semiconductor are in a favourable position to catalyse the solar splitting of water under a given set of conditions. In the case of carbon dioxide conversion to fuels (e.g. methane, $CH_4$, methanol, $CH_3OH$, or formic acid, $HCO_2H$), the position of the conduction band of the semiconductor photocatalyst has to be above the redox potential for the $CO_2$ reduction half-reaction, which depends on the specific fuel produced. Since the $CO_2/CH_4$, $CO_2/CH_3OH$ and $CO_2/HCO_2H$ levels are above the HER ($H^+/H_2$) level, the photocatalyst for the $CO_2$ reduction reactions requires a minimum bandgap that is wider than for water splitting.

Figure 5 shows the bandgaps and band edge positions of the six material compositions studied in this work, with respect to the vacuum level, as calculated using the HSE06 functional. The values of the redox potentials (at neutral pH and room temperature) for the species appearing in the water splitting reaction and for the reactions in which carbon dioxide is reduced to produce methane, methanol, and formic acid, are also shown in the figure: $E(H_2O/O_2)$=-5.26 eV; $E(H^+/H_2)$=-4.03 eV; $E(CO_2/CH_4)$=-3.79 eV; $E(CO_2/CH_3OH)$=-3.65 eV; $E(CO_2/HCOOH)$=-3.42 eV.[88]

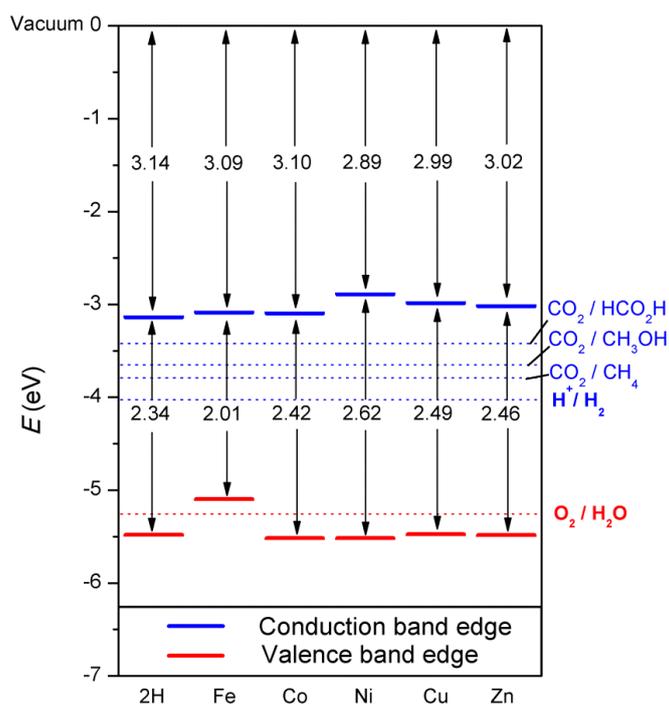

Figure 5. Bandgaps and band edge positions with respect to the vacuum level, as calculated with the HSE06 functional. Energy levels corresponding to redox potentials of water splitting and carbon dioxide reduction reactions producing methane, methanol, and formic acid at pH = 7 are also shown with dotted lines.

For all materials, the conduction band edge is at roughly the same position (ca. -3 eV in the vacuum scale), which is above the energies corresponding to the $H^+/H_2$ and $CO_2/CH_xO_y$ levels. That means that the MOF would be thermodynamically able to donate an excited electron from the conduction band for the reduction half-reactions to proceed. The conduction bands are slightly higher than desired for water splitting, but they are in a nearly ideal position for the carbon dioxide reduction reactions.

As for the oxygen evolution half-reaction, in all materials, except Fe-Al-PMOF, the valence band edge lies at about -5.5 eV in the vacuum scale, below the energy of the $O_2/H_2O$ level. In the case of Fe-Al-PMOF, it lies slightly (~0.2 eV) above. Of course, due to the approximations made in our calculations, such small energy differences are not reliable. But it is clear that the valence band edge for all compositions is in the correct energy range in the absolute scale. Small deviations from the ideal band edge positions in a semiconductor with the adequate bandgap can always be corrected via the application of a weak bias voltage in a photoelectrochemical device, shifting both band edges with respect to the redox levels (applying a voltage increases device complexity and also consumes energy, so a large bias voltage should be avoided). For example, in a recent first-principles screening of materials for water splitting photocatalysts, Wu et al. set the threshold of the allowed bias voltage to 0.7 V.[89] Our present calculations show that the Al-PMOF, in all the different compositions explored here, would be able to operate as single-semiconductor photocatalyst with little or no bias voltage applied at neutral pH.

## Conclusions

We have carried out a theoretical study of the electronic properties of a porphyrin-based metal organic framework, including both protonated and metallated ($Fe^{2+}$, $Co^{2+}$, $Ni^{2+}$, $Cu^{2+}$, $Zn^{2+}$) porphyrins. We have found that the protons would be spontaneously exchanged with these cations, when the materials are submerged in aqueous solutions of these cations. The analysis of the electronic bands reveals that the bandgaps of all materials are in the favourable range for efficient adsorption of solar light (2.0 to 2.6 eV). Furthermore, the alignment of the bands is also favourable in all cases for the photocatalysis of water splitting and carbon dioxide reduction, which means that a device using these materials would require little or no bias voltage to function. Our calculations only show small variations in the electronic band edges of the metal-substituted structures. The only exception is the Fe-substituted one, where an occupied $d_{xy}$ state is introduced in the gap region above the valence band. In all the other cases studied here, the band edges are determined by the porphyrin electronic structure. This means that the choice of metal at the porphyrin centre could be used to optimise other properties, e.g. molecular adsorption, affecting the photocatalytic process. Our calculations demonstrate that these porphyrin-based MOFs are very promising candidates for efficient photocatalysis of fuel production reactions. Further studies about the adsorption and diffusion of molecules in these structures would be helpful to select the best composition for performing the photocatalytic reactions studied.

## Acknowledgements

We are grateful to the Royal Society for an International Exchange Scheme grant. Via our membership of the UK's HPC Materials Chemistry Consortium, which is funded by EPSRC (EP/L000202), this work made use of the facilities of ARCHER, the UK's national high-performance computing services, which are funded by the Office of Science and Technology through EPSRC's High End Computing Programme. This work was also supported by the European Research Council through an ERC Starting Grant (ERC2011-StG-279520-RASPA), by the MINECO (CTQ2013-48396-P) and by the Andalucía Region (FQM-1851).

## Notes and references

## Supplementary Information

**Effects of Hubbard and dispersion corrections on the results**

In our simulations, we used the GGA-PBE functional to optimise the geometries, while the electronic structures were calculated using single-point runs with the HSE06 functional. We have followed this approach because we are dealing with large simulation cells, where it becomes very computationally expensive to carry out the optimisations at the HSE06 level.

Here we provide evidence to show that the errors in the PBE optimisation are not expected to carry over to the electronic structure calculation. We have selected one of the structures (the one with Ni, which in this case is non spin-polarised, and therefore easier to calculate) for the test calculations.

In Table S1, we compare the geometries obtained with the following methods: a) PBE, b) PBE-D2, where the van der Waals interactions are taken into account via the DFT-D2 method of Grimme, c) PBE+U, where the $U_{eff}$ parameter that describe the on-site Coulomb interaction of d electrons is set at 3 eV, and d) PBE-D2+U (i.e., both corrections applied).

Table S1. Calculated lattice parameters, cell volume and the two perpendicular N-N distances inside the porphyrin for the Ni-Al-PMOF structure, using different DFT functionals for the optimisation. As in Table 1 of the manuscript, in all cases the structures adopt the orthorhombic space group *Cmmm* (65), where α=β=γ=90°.

| Functional | *a* (Å) | *b* (Å) | *c* (Å) | *V* (Å3) | *d* [N-N] (Å) |
|---|---|---|---|---|---|
| PBE | 31.995 | 6.723 | 16.964 | 3649.2 | 3.91/3.91 |
| PBE-D2 | 31.920 | 6.641 | 16.949 | 3592.6 | 3.91/3.91 |
| PBE+U | 31.999 | 6.722 | 16.969 | 3649.8 | 3.92/3.93 |
| PBE-D2 + U | 31.917 | 6.644 | 16.956 | 3595.4 | 3.92/3.92 |

The cell parameters exhibit little deviation from one calculation to the other (maximum discrepancy is ~1% along the *b* axis). The small effect of the U correction on the cell parameters is due to the fact that it only affects the metal at the centre of the rigid porphyrin ligand. On the other hand, the effect of the D2 correction is also small (although larger than for the U correction) because the whole structure is covalently linked.

The question remains on whether these small geometric changes can affect the final electronic structure. Figure S1 shows the positions of the band edges, as calculated with an HSE06 single point on the geometries optimised with different functionals. Clearly the variations in the positions of the band edges are very small (less than 0.04 eV).

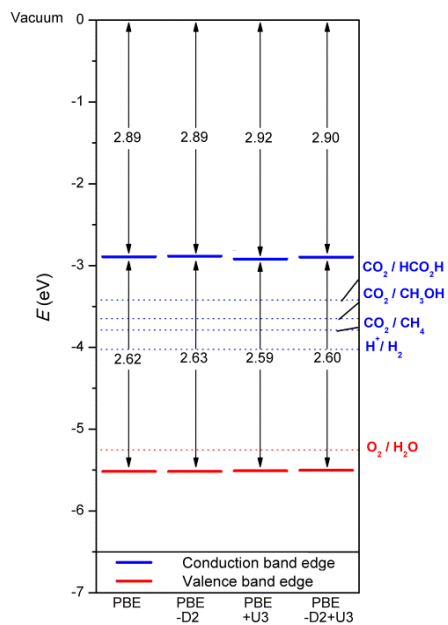

Figure S1. Bandgaps and band edge positions of the Ni-Al-PMOF system with respect to the vacuum level, as calculated with the HSE06 functional on geometries optimised at different levels of calculations (PBE, PBE-D2, PBE+U with U=3 eV, and PBE-D2+U). Energy levels corresponding to redox potentials of water splitting and carbon dioxide reduction reactions producing methane, methanol, and formic acid at pH = 7 are also shown with dotted lines.